# Teaching a Language Model to Speak the Language of Tools

## Simeon Emanuilov

*Department of Software Technologies, Faculty of Mathematics and Informatics, Sofia University "St. Kliment Ohridski"*
*Email: ssemanuilo@fmi.uni-sofia.bg*

External tool integration through function-calling is essential for practical language model applications, yet most multilingual models lack reliable tool-use capabilities in non-English languages. Even state-of-the-art multilingual models struggle with determining when to use tools and generating the structured outputs required for function calls, often exhibiting language confusion when prompted in lower-resource languages. This work presents a methodology for adapting existing language models to enable robust tool use in any target language, using Bulgarian as a case study. The approach involves continued training of the BgGPT model series (2.6B, 9B, 27B parameters) on a novel bilingual dataset of 10,035 function-calling examples designed to support standardized protocols like MCP (Model Context Protocol). The research introduces TUCAN (Tool-Using Capable Assistant Navigator), which achieves up to 28.75% improvement in function-calling accuracy over base models while preserving core language understanding, as verified on established Bulgarian benchmarks. Beyond accuracy gains, TUCAN models demonstrate production-ready response formatting with clean, parsable function calls, contrasting with the verbose and inconsistent outputs of base models. The models, evaluation framework, and dataset are released to enable replication for other languages. This work demonstrates a practical approach for extending tool-augmented capabilities beyond English-centric systems.

## 1. Introduction

The emergence of language models has fundamentally transformed how we interact with artificial intelligence, with open-source developments rapidly approaching the capabilities of proprietary systems (Zhang, et al. 2024). However, the true transformative potential of these models extends far beyond their inherent linguistic competencies—it lies in their ability to transcend static knowledge boundaries and orchestrate dynamic interactions with external systems. This capability, achieved through function-calling or tool use, elevates language models from text generators to autonomous agents capable of executing real-world tasks through database queries, API interactions, and software integration.

Despite these advances, the benefits remain disproportionately concentrated within English-centric ecosystems (Zhao, et al. 2024) (Joshi, et al. 2024). A critical capability gap persists across multilingual contexts, where even models with strong comprehension abilities falter when tasked with tool integration. Consider state-of-the-art Bulgarian language models like BgGPT (Alexandrov, et al. 2024): while demonstrating exceptional language understanding, they require additional specialization for reliable tool use. This limitation manifests as failures to invoke functions when required, generation of incorrect or hallucinated parameters, and explicit discussion of function mechanics rather than proper execution. Such barriers significantly constrain the development of AI applications for global audiences.

Traditional approaches to inducing specialized behaviors from language models often rely heavily on prompt engineering. However, the exacting syntactic requirements of function-calling render such approaches insufficient for consistent performance (Geng, et al. 2025). Even with capable foundation models and carefully designed system prompts, output reliability remains problematic—a critical bottleneck given that AI applications require predictable, machine-parsable responses for correct operation. This formatting instability underscores the inadequacy of prompt engineering as a standalone solution, highlighting the need for more robust approaches such as specialized fine-tuning (Scialom, Chakrabarty and Muresan 2022).

Moreover, effective tool integration demands capabilities beyond simple function triggering. A truly capable model must orchestrate complete conversational workflows around tool usage: intelligently requesting clarification for incomplete user inputs, discerning when to leverage tools versus internal knowledge, and managing the complex interaction patterns that define AI assistants. For instance, when a user requests "book a flight to Barcelona" an effective system should clarify missing details (dates, preferences) before calling booking APIs, rather than failing or making inappropriate assumptions. The Model Context Protocol (MCP) represents an emerging standardization effort for these multi-turn interactions between users, models, and external tools, providing structured frameworks for next-generation AI agent development (Hou, et al. 2025).

This paper addresses these multilingual tool-use challenges by presenting a systematic methodology for adapting language-specific models for function-calling capabilities. The research introduces TUCAN (Tool-Using Capable Assistant Navigator), a series of open-source Bulgarian models specifically fine-tuned for function-calling through continued training of state-of-the-art BgGPT models on a novel bilingual dataset exceeding 10,035 function-calling examples. The approach demonstrates substantial improvements in tool-use capability—achieving up to 28.75% accuracy gains in function-calling—while avoiding catastrophic forgetting as verified on Bulgarian variants of established benchmarks, including HellaSwag (Zellers, et al. 2019), Winograde (Sakaguchi, et al. 2021), and ARC-Challenge (Clark, et al. 2018). The improvements extend across all model sizes (2.6B, 9B, 27B), with enhanced conversational flow and contextual understanding enabling more natural multi-turn interactions around tool usage.

The complete TUCAN model series is released alongside a dedicated evaluation framework and the training dataset. This work establishes a replicable blueprint for extending tool-augmented capabilities beyond English to serve global linguistic communities..

## 2. Related Work

The intersection of multilingual language models and tool use capabilities represents a rapidly evolving research landscape. This section examines the foundational developments in multilingual model adaptation, tool use frameworks, and standardization protocols that enable our work.

### 2.1 Multilingual Language Models and Low-Resource Language Adaptation

Contemporary foundational models such as Google's Gemma family (Team, Riviere, et al. 2024) (Team, Kamath, et al. 2025), Meta's Llama series (Grattafiori, et al. 2024), and Alibaba's Qwen models (Yang, et al. 2025) demonstrate remarkable multilingual capabilities while serving as a starting point for language-specific adaptations.

A critical issue in multilingual modeling is addressing the "curse of multilinguality" (Conneau, et al. 2019), where massively multilingual training dilutes performance for individual languages. Recent work has focused on targeted adaptation approaches: EMMA-500 (Ji, et al. 2024) and MaLA-500 (Lin, et al. 2024) extend language coverage to 500+ languages through continual pre-training, while studies show that targeted multilingual adaptation can outperform both monolingual and massively multilingual baselines for related language families (Downey, et al. 2024).

The Bulgarian language modeling shows successful language-specific adaptation through the BgGPT series (Alexandrov, et al. 2024), which employs the "Branch-and-Merge" technique during continued pretraining to preserve foundational capabilities while integrating Bulgarian-specific knowledge. This methodology provides a direct precedent for our approach.

## 2.2 Function-Calling and Tool Use in Language Models

The evolution from text generators to autonomous agents capable of external tool integration represents a fundamental shift in AI capabilities. Early systems like ToolAlpaca (Tang, et al. 2023) and ToolLLM (Qin, et al. 2023) established approaches for API integration and simple tool use scenarios.

Recent work has focused on complex tool interaction patterns. The Berkeley Function-Calling Leaderboard (BFCL) (Yan, et al. 2024) has emerged as the leading benchmark for evaluating function-calling capabilities, while ToolSandbox (Lu, et al. 2024) introduces stateful, conversational evaluation scenarios that better reflect real-world deployment conditions. However, both frameworks remain predominantly English-centric.

Contemporary research addresses critical challenges, including determining when tools are necessary versus relying on internal knowledge (MetaTool benchmark) (Huang, et al. 2023), handling multi-turn conversational flows, and managing precise parameter extraction for reliable function execution. UltraTool (Huang, et al. 2024) evaluates the entire tool utilization pipeline from planning through execution, while T-Eval (Chen, et al. 2023) provides a step-by-step assessment of tool utilization capabilities.

## 2.3 Parameter-Efficient Fine-Tuning and Evaluation Frameworks

The computational demands of full fine-tuning have driven innovation in parameter-efficient adaptation methods. Low-Rank Adaptation (LoRA) (Hu, et al. 2022) enables substantial capability improvements while updating only a small fraction of model parameters. Recent work demonstrates that soft-prompt tuning can achieve effective cross-lingual transfer using merely 0.28% of tuned parameters (Chen and Chen 2024).

The challenge of catastrophic forgetting has been addressed through targeted domain-specific adaptations that preserve general capabilities while adding new competencies (Shi, et al. 2024). For Bulgarian specifically, evaluation relies on translated benchmarks[1], including HellaSwagBG, WinograndeBG, and ARC Challenge, assessed using a clone of the original *lm-evaluation-harness* framework (Gao, et al. 2024).

## 2.4 Standardization and Research Gaps

The Model Context Protocol (MCP) (Hou, et al. 2025) represents an emerging standardization effort defining structured frameworks for multi-turn interactions between users, AI models, and external tools, providing interfaces for tool discovery, context management, and conversation history tracking.

---

[1] Evaluated using the lm-evaluation-harness-bg framework (INSAIT Institute): https://github.com/insait-institute/lm-evaluation-harness-bg/

Despite significant progress, critical issues remain: most of the function-calling research remains English-centric, existing multilingual adaptation techniques have not been systematically evaluated for tool use capabilities, and evaluation frameworks for assessing both linguistic competence and functional tool integration across multiple languages are underexplored.

This work addresses these limitations by demonstrating a systematic methodology for adapting language-specific models for function-calling, introducing an evaluation framework tailored for multilingual tool use assessment, and providing empirical evidence that specialized fine-tuning can enhance tool capabilities without compromising core language understanding.

# 3. Dataset

Effective function-calling capabilities require training data that goes far beyond standard instruction-following datasets. Traditional datasets often lack the conversational complexity and structural precision necessary for robust tool use. To address this, a bilingual fine-tuning dataset of 10,035 conversations was developed, specifically engineered to teach models not only *when* to call functions but also *when not to*, and how to navigate the intricate conversational dynamics that surround tool interaction.

## 3.1. Generation Methodology

Dataset creation followed a systematic hybrid approach combining manual curation with synthetic generation. High-quality "gold standard" examples across diverse function-calling scenarios were manually created, serving as few-shot exemplars for generative models including OpenAI's GPT-4.1[2] (40%), Google's Gemini 2.5 Pro[3] (30%), and Anthropic's Claude Sonnet 4[4] (30%).

The bilingual nature of the dataset reflects realistic deployment scenarios: function definitions and parameters are specified mostly in English (following standard developer practices), while conversational exchanges between users and assistants occur in Bulgarian.

## 3.2. Dataset Structure and Tag System

**JSON-Line Format and Conversation Structure:** The dataset employs JSON-Line (NDJSON) format, with each line representing a complete conversation structured as individual messages containing "from" and "value" fields. The "from" field identifies message senders ("user", "model", or "tool"), while "value" contains message content.

**Tool Definition System:** Tool definitions are embedded within user messages using XML-style tags. The **<tools>** tag encapsulates JSON arrays of function definitions, each including "name", "description", and "parameters" schema.

**Function Call Execution Tags:** Function execution is indicated through **<tool_code>** tags containing function names and arguments in JSON format.

---

[2] GPT-4.1 models from OpenAI, https://openai.com/index/gpt-4-1/
[3] Gemini 2.5 Pro from Google, https://deepmind.google/models/gemini/pro/
[4] Claude Sonnet 4 from Anthropic, https://www.anthropic.com/claude/sonnet

**Tool Response Integration:** Tool responses are wrapped in **<tool_response>** tags, providing structured execution feedback. This enables understanding of function call outcomes and generation of appropriate follow-up responses.

The tag system serves multiple purposes: **<tools>** provides function context and capabilities, **<tool_code>** enables structured function execution with proper parameter formatting, and **<tool_response>** delivers execution results in a parsable format. Empty **[]** in tools signals contexts where no tools are available, distinguishing between tool-using and non-tool-using scenarios.

The tag format serves as a structural template that can be adapted to specific model architectures and prompt formatting requirements during training.

Figure 1 illustrates four representative scenario types from the dataset, demonstrating the range of multi-turn interactions models must handle: direct function execution, parameter clarification workflows, answer from internal knowledge or appropriate tool declination.

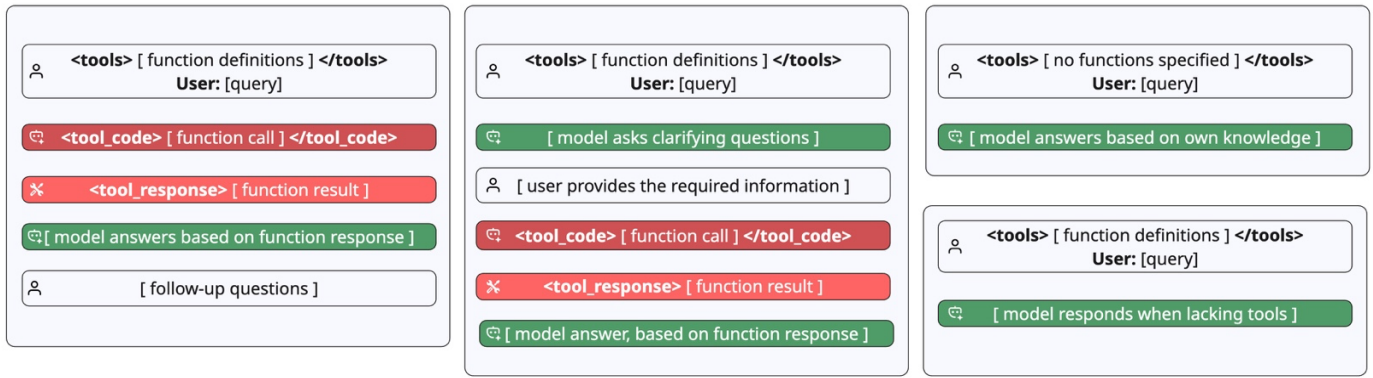

**Figure 1. Representative conversation flows in the TUCAN training dataset showing (left) direct function call execution, (center) multi-turn clarification before tool use, and (right) appropriate tool declination when functions are unavailable or inappropriate for the user query.**

## 3.3. Conversational Structure and Complexity

The dataset exhibits rich conversational patterns capturing real-world interaction complexity. Conversations range from 1 to 15 messages (mean: 4.4, median: 4.0).

**Table 1. Conversation Structure Statistics**

| Metric | Min | Max | Mean | Median |
|---|---|---|---|---|
| Messages per conversation | 1 | 15 | 4.4 | 4.0 |
| User messages per conversation | 1 | 6 | 1.4 | 1.0 |
| Tool messages per conversation | 0 | 4 | 0.8 | 1.0 |

## 3.4. Model Behavior Analysis and Overlapping Patterns

Behavioral metrics represent overlapping conversation characteristics rather than disjoint categories, explaining why percentages exceed 100% when summed.

**Tool Usage (72.57%):** Conversations where successful function execution occurs, indicating successful identification and execution of function calls in most conversations.

**Rejection Behavior (16.54%):** Conversations with explicit tool declination when available functions are inappropriate for requests. This demonstrates robust rejection mechanisms preventing inappropriate tool usage.

**Clarification Requests (67.69%):** Conversations requesting additional information before proceeding.

These behavioral characteristics can co-occur within single conversations. For instance, a conversation might begin with clarification requests, then proceed to tool calls.

## 3.5. Function Diversity and Domain Coverage

The dataset encompasses 16,097 function definitions spanning approximately 2,204 distinct topics[5], ensuring broad domain coverage. Function descriptions are predominantly English (86.7%), with Bulgarian descriptions (11.5%) and mixed-language descriptions (1.8%) providing additional linguistic diversity.

**Table 2. Function Language Distribution**

| Language | Count | Percentage | Example Functions |
|----------|-------|------------|-------------------|
| English | 13,957 | 86.7% | `run_applescript`, `generate_backup` |
| Bulgarian | 1,857 | 11.5% | `generate_digital_signature`, `verify_digital_signature` |
| Mixed | 283 | 1.8% | `get_national_olympiad_info` |

The most frequent function patterns reflect common developer and user scenarios, with retrieval operations (*get_, list_, find_*) dominating distribution, followed by creation and manipulation functions.

## 3.6. Topic Distribution

To assess domain coverage breadth, automated topic classification across all 2,204 distinct topics was performed using GPT-4o-mini. Classification employed hierarchical categorization designed to capture both technical and everyday use cases, reflecting diverse scenarios where function-calling capabilities are needed.

**Personal & Lifestyle (17.9%)** represents the largest category, including health tracking, home automation, and travel planning. **Government & Services (16.8%)** encompasses document processing, civic services, and legal workflows. **Business & Finance (12.2%)** covers payment processing, CRM, and financial analysis. **Specialized & Niche (10.7%)** includes industry-specific tools and emerging technologies. **Technology & Development (8.9%)** encompasses API integration, software deployment, and debugging. **Data & Analytics (8.3%)** covers business intelligence, reporting, and visualization. **Content & Media (7.4%)** includes content creation, media processing, and publishing. **Infrastructure & DevOps (6.8%)** covers cloud services, monitoring, and system administration. **Education & Learning (6.4%)** encompasses learning platforms, academic tools, and training. **Communication & Social (4.7%)** includes messaging, collaboration, and social media.

Figure 2 shows the topic distribution in the different categories.

---

[5] Complete topic list available at: https://huggingface.co/datasets/llm-bg/Tucan-BG-v1.0/blob/main/topics.txt

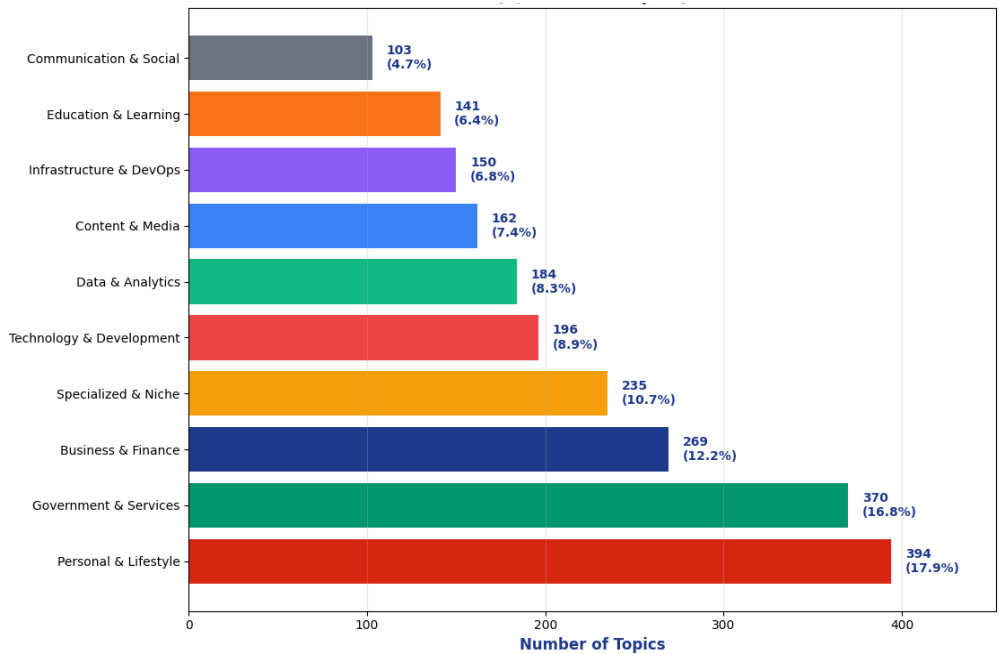

**Figure 2. Topic Distribution Across Major Categories**

This topic coverage ensures real-world scenario exposure during training, from everyday personal tasks to specialized professional workflows. The balanced distribution prevents overfitting to specific domains while ensuring adequate representation across all major use case categories.

## 3.6. Message Length and Content Analysis

The dataset exhibits realistic message length distributions reflecting natural conversation patterns:

- User messages: 545 characters average (range: 5-3,683)
- Model responses: 340 characters average (range: 16-3,720)
- Tool responses: 246 characters average (range: 43-1,941)

These length distributions ensure learning to handle both concise queries and detailed explanations while maintaining appropriate response verbosity.

The dataset design enables function-calling capabilities while maintaining natural conversational flow and cross-linguistic competence.

## 4. Model

The TUCAN model series represents adaptations of existing base models rather than training from scratch. This approach leverages pre-training and reasoning capabilities of established architectures while adding function-calling capabilities through targeted fine-tuning.

## 4.1 Base Model Selection and Architecture

The foundation utilizes the Bulgarian BgGPT model family, specifically the BgGPT-Gemma-2 variants developed by the INSAIT Institute[6]. Three parameter scales are employed: 2.6B, 9B, and 27B parameters, all built upon Google's Gemma-2 architecture. These models provide the multilingual capabilities and reasoning abilities necessary for tool use in Bulgarian language contexts.

The model's primary responsibilities include: (1) determining when a tool is required to fulfill a user's request, (2) identifying the correct function from available tools, (3) generating precise tool calling objects with appropriate arguments, and (4) formulating natural language responses based on tool execution results.

## 4.2 Parameter-Efficient Fine-Tuning Methodology

To adapt the base BgGPT models for function-calling while preserving their existing knowledge and general reasoning capabilities, Parameter-Efficient Fine-Tuning (PEFT) using Low-Rank Adaptation (LoRA) was employed. This approach was selected over full fine-tuning for computational efficiency and catastrophic forgetting prevention. By updating only a small subset of parameters, LoRA reduces computational requirements while preserving the base models' existing capabilities.

To enhance practical deployment, 4-bit quantization during loading (`load_in_4bit = True`) was implemented, reducing memory footprint and enabling execution on accessible hardware without significant performance degradation. The fine-tuning process utilized the Unsloth library[7] for optimized training implementations.

## 4.3 Structured Prompt Template and Format

A critical component is the structured prompt template, enforced during both fine-tuning and inference to ensure reliable function-calling behavior. This template guides the model through role definition, tool usage instructions, function schema integration, and query delineation in Gemma's appropriate style. The dataset was formatted to convert the XML placeholders into the structure below:

```
<bos><start_of_turn>user
Ти си полезен AI асистент, който предоставя полезни и точни отговори.
Имаш достъп и можеш да извикаш една или повече функции, за да помогнеш с
потребителското запитване.
Използвай ги, само ако е необходимо и подходящо.
Когато използваш функция, форматирай извикването й в блок ```tool_call``` на отделен
ред, а след това ще получиш резултат от изпълнението в блок ```toll_response```.
## Шаблон за извикване:
```tool_call
{"name": <function-name>, "arguments": <args-json-object>}```
## Налични функции:
[your function definitions in JSON format here]
## Потребителска заявка:
[your query in Bulgarian]<end_of_turn>
<start_of_turn>model
```

---







## 4.3 Hyperparameter Configuration and Training Statistics

The models were fine-tuned with consistent hyperparameters across different sizes to ensure controlled comparison. Training was performed on NVIDIA L40S GPU for the 27B model and NVIDIA L4 GPUs for the 9B and 2.6B models. All models used eager attention implementation (required for Gemma-2 architecture), cosine learning rate scheduling with 0.1 warmup ratio, and adamw_8bit optimizer with 0.01 weight decay for regularization.

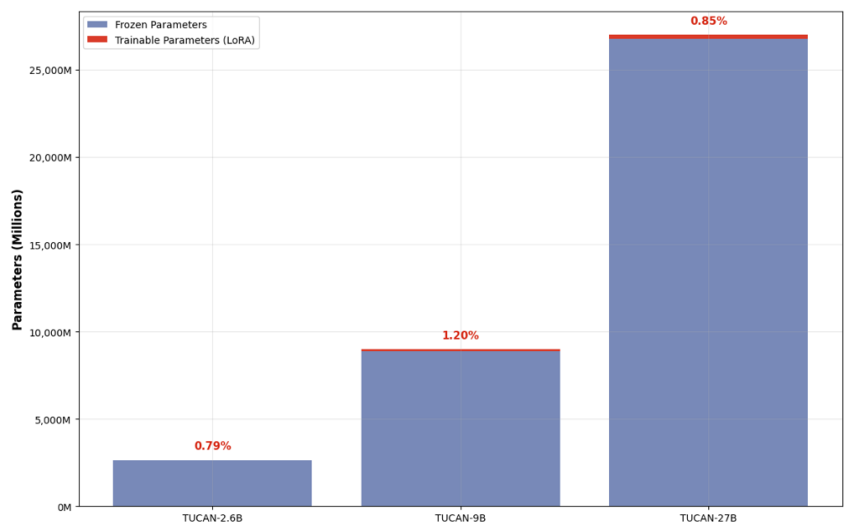

**Figure 3. LoRA adaptation strategy**

**Table 3. Hyperparameter Configuration for TUCAN Model Training**

| Parameter | 2.6B Model | 9B Model | 27B Model |
|---|---|---|---|
| Base | BgGPT-2.6B-IT-v1.0 | BgGPT-9B-IT-v1.0 | BgGPT-27B-IT-v1.0 |
| LoRA Rank (r) | 16 | 32 | 32 |
| LoRA Alpha | 16 | 32 | 32 |
| Trainable parameters | 20,766,720 | 108,036,096 | 228,360,192 |
| % trained | 0.79 | 1.2 | 0.85 |
| Epochs | 3 | 3 | 3 |

Regarding LoRA rank selection, empirical experimentation revealed that the 2.6B model exhibited increased instability and performance degradation when higher LoRA rank and alpha values were employed. This manifested as erratic loss curves and occasional breakdowns in function-calling behavior. Consequently, a conservative configuration (rank and alpha of 16) was adopted for the 2.6B model to ensure stable convergence. In contrast, the larger 9B and 27B models demonstrated better training dynamics with higher LoRA ranks (32), allowing for greater adaptation capacity without compromising stability.

## 5. Evaluations

To assess the effectiveness of the fine-tuning methodology, a two-part evaluation was conducted. First, a specialized framework was developed to measure function-calling accuracy in multilingual contexts. Second, knowledge retention was evaluated on standard Bulgarian language benchmarks to verify that enhanced tool-use capabilities did not compromise core linguistic competence.





## 5.1 Function-Calling Evaluation Framework

The Tucan-Eval framework (https://github.com/llm-bg/Tucan-Eval) was developed and open-sourced as a unified command-line interface for evaluating language models on function-calling tasks. This framework represents a key contribution of the work, providing a standardized evaluation methodology for multilingual tool-use assessment.

### 5.1.1 Framework Architecture

The Tucan-Eval framework employs a CLI-first design that eliminates configuration file requirements while supporting multiple model architectures. Key features include:

- **Zero-configuration setup** for BgGPT/Gemma models with optimized defaults (bfloat16 precision, eager attention implementation, 4-bit quantization)
- **Multi-platform support** for Hugging Face transformers (Wolf, et al. 2020), OpenAI API, and local models
- **Multilingual capability** with customizable prompts, headers, and prefixes
- **Comprehensive analysis** providing detailed accuracy metrics and error distribution

### 5.1.2 Evaluation Methodology

The framework implements a systematic four-stage evaluation process:

1. **Tool Call Parsing:** The system extracts JSON objects enclosed within predefined tags (`tool_call` blocks) from model-generated responses. Malformed JSON triggers `MALFORMED_JSON` error classification.
2. **Behavioral Validation:** Parsed outputs are compared against expected behaviors defined in the evaluation dataset. The framework verifies whether function calls were appropriately made or avoided based on scenario requirements.
3. **Function Validation:** For scenarios requiring function calls, the framework validates both function name selection against `expected_function` and parameter accuracy against `expected_parameters`.
4. **Parameter Comparison:** Arguments undergo robust validation with lenient type matching, string-to-number coercion, and normalization for case, punctuation, and Cyrillic-to-Latin transliteration differences.

### 5.1.3 Error Classification System

The framework categorizes five primary error types that capture distinct failure modes in function-calling behavior:

- **NO_CALL_WHEN_EXPECTED:** Models fail to generate function calls when tools are required to fulfill user requests
- **UNEXPECTED_CALL:** Models generate inappropriate function calls for queries that should be answered from internal knowledge or when no suitable tools are available
- **WRONG_FUNCTION:** Models correctly identify the need for tool use but select inappropriate functions from the available set





- **WRONG_PARAMETERS:** Models select correct functions but generate incorrect, incomplete, or malformed argument structures
- **MALFORMED_JSON:** Models produce syntactically invalid JSON structures that cannot be parsed for function execution

### 5.1.4 Scenario-Based Assessment

The evaluation employs six distinct scenario types designed to assess function-calling capabilities:

1. **Function Call Required:** Scenarios where user queries explicitly require tool use for task completion, testing models' ability to recognize tool necessity and execute appropriate calls.
2. **Multiple Functions Selection:** Complex scenarios presenting multiple available tools, requiring models to demonstrate sophisticated reasoning about function appropriateness and context matching.
3. **Irrelevant Question with Functions:** Test cases where functions are available but user queries require general knowledge responses, assessing models' ability to avoid unnecessary tool use.
4. **No Functions Available:** Scenarios without available tools, verifying models' capability to provide text-only responses when function-calling is impossible.
5. **Ambiguous Function Selection:** Cases where multiple functions could potentially address user requests, requiring nuanced reasoning about optimal tool selection.
6. **Missing Required Parameters:** Scenarios where user queries lack complete information needed for function execution, testing models' ability to request clarification or handle incomplete inputs appropriately.

### 5.1.5 Evaluation Dataset and Metrics

Function-calling assessment utilizes the Tucan-BG-Eval-v1.0 dataset (https://huggingface.co/datasets/llm-bg/Tucan-BG-Eval-v1.0), containing 120 carefully curated test cases distributed equally across the six scenario types (20 cases each). The dataset covers diverse domains including government services, business applications, personal tasks, and technical operations, ensuring comprehensive evaluation across realistic use cases.

The primary evaluation metric is overall accuracy, calculated as the percentage of test cases where models demonstrate completely correct behavior according to scenario requirements. Additional metrics include scenario-specific accuracy rates and detailed error distribution analysis, providing granular insights into model strengths and failure patterns.

## 5.2 Knowledge Retention Evaluation

To verify that function-calling specialization did not induce catastrophic forgetting, TUCAN models were evaluated against original BgGPT models using the lm-evaluation-harness-bg repository[8] (a clone of the original evaluation harness[9]).

---

[8] Language Model Evaluation Harness with Bulgarian benchmarks, https://github.com/insait-institute/lm-evaluation-harness-bg/

[9] A framework for few-shot evaluation of language models, https://github.com/EleutherAI/lm-evaluation-harness





5.2.1 Benchmark Suite

The models were assessed on four established Bulgarian language benchmarks:

- **HellaSwagBG:** Commonsense reasoning through sentence completion tasks
- **WinograndeBG:** Coreference resolution requiring world knowledge and contextual understanding
- **ARC-Easy-BG** and **ARC-Challenge-BG**: Science question answering with varying difficulty levels, testing factual knowledge and reasoning capabilities

5.2.2 Evaluation Protocol

All models were evaluated using recommended settings from BgGPT documentation and Gemma 2 specifications to ensure fair comparison. Identical quantization settings and generation parameters were employed across all models to maintain evaluation consistency. The evaluation protocol followed standard practices for the lm-evaluation-harness framework, including few-shot prompting where appropriate and standardized metric computation.

5.2.3 Comparative Analysis Framework

The evaluation objective was knowledge retention verification rather than state-of-the-art performance achievement, which would require full-precision weights and extensive hyperparameter optimization beyond the scope of this work. TUCAN models were compared directly against their corresponding BgGPT base models to assess whether fine-tuning preserved core linguistic competencies while adding function-calling capabilities.

The evaluation framework and methodology provide reproducible assessment tools for future research in multilingual tool-use capabilities. They establish standardized metrics for comparing function-calling performance across different languages and model architectures.

# 6. Results

All performance metrics were obtained by averaging results across multiple independent evaluation runs to ensure statistical reliability. Each model was evaluated several times on the Tucan-BG-Eval-v1.0 dataset for function-calling assessment, with final scores representing the mean accuracy across runs. Knowledge retention evaluation on Bulgarian benchmarks followed the same multi-run protocol.

## 6.1 Function-Calling Performance

The primary objective of this work was to enhance function-calling capabilities in Bulgarian language models. Figure 5 presents the function-calling accuracy results measured using our Tucan-Eval framework on 120 test cases spanning six distinct scenario types.





**Table 3. Function-Calling Accuracy Results (Averaged Across Multiple Runs)**

| Model | Overall Accuracy | Improvement |
|---|---|---|
| BgGPT-Gemma-2-2.6B-IT-v1.0 | 50.00% | - |
| Tucan-2.6B-v1.0 | **78.75%** | **+28.75%** |
| BgGPT-Gemma-2-9B-IT-v1.0 | 78.33% | - |
| Tucan-9B-v1.0 | **86.67%** | **+8.34%** |
| BgGPT-Gemma-2-27B-IT-v1.0 | 86.67% | - |
| Tucan-27B-v1.0 | **87.50%** | **+0.83%** |

The figure below visualizes the performance gain of Tucan versus BgGPT models.

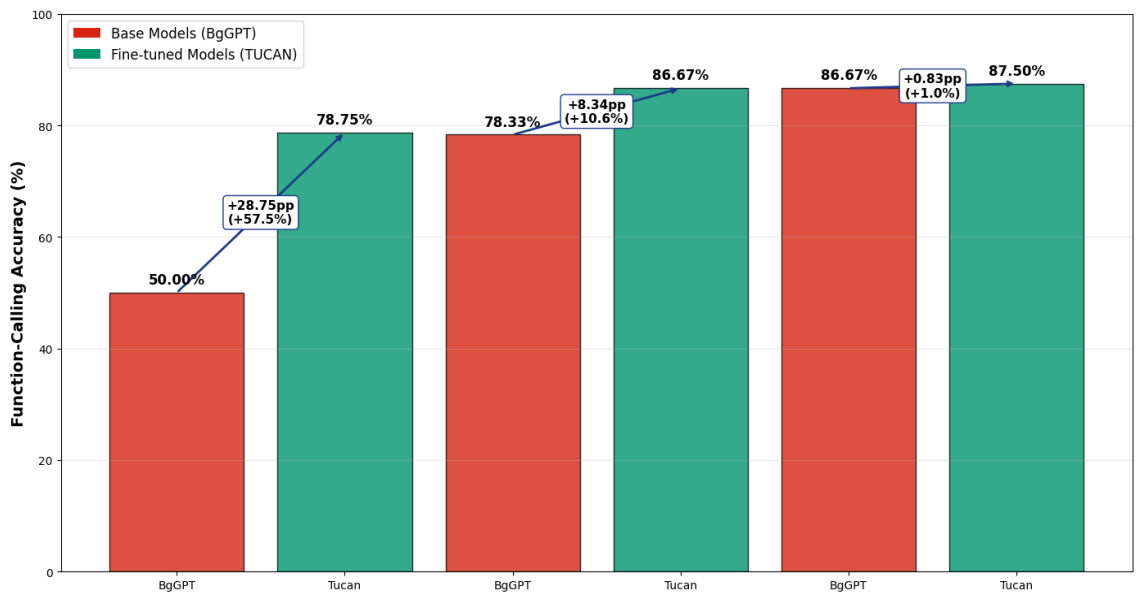

**Figure 4. Tucan model performance, compared to BgGPT**

The TUCAN models demonstrate improvements across all parameter scales. The most significant gains occur in smaller models, with **Tucan-2.6B** achieving a 57.5% relative improvement (28.75 percentage points) over its base model. The **Tucan-9B** model shows a 10.7% relative improvement (8.34 percentage points), while **Tucan-27B** achieves a 1.0% relative improvement (0.83 percentage points). All improvements represent enhancements in function-calling capability and response quality.

## 6.2 Scenario-Specific Performance Analysis

Figure 6 presents performance breakdown across the six evaluation scenarios, revealing distinct patterns in model behavior and highlighting the critical improvements achieved by TUCAN models.





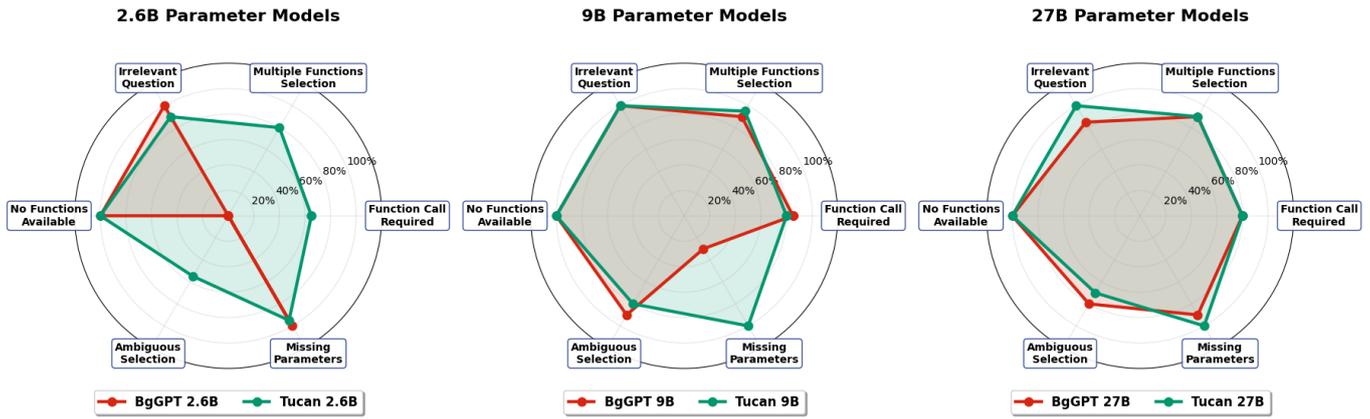

**Figure 5. Tucan vs BgGPT for specific tasks**

The most critical finding is the complete failure of **BgGPT-2.6B** in multiple key scenarios: "Function Call Required" (0% accuracy), "Multiple Functions Selection" (0% accuracy), and "Ambiguous Selection" (0% accuracy), where the model consistently failed to recognize when and which tools were necessary. In contrast, **Tucan-2.6B** achieved 65% accuracy in "Function Call Required", 80% in "Multiple Functions Selection", and 55% in "Ambiguous Selection", representing fundamental improvements in tool-use recognition.

TUCAN models demonstrate superior and consistent performance in parameter handling scenarios. In "Missing Required Parameters" scenarios, **Tucan-9B** and **Tucan-27B** achieve perfect accuracy (100%), while **Tucan-2.6B** achieves 95% accuracy. **BgGPT-2.6B** also achieves perfect accuracy (100%) in this scenario, but **BgGPT-9B** drops dramatically to only 30% accuracy, suggesting inconsistent scaling behavior in the BgGPT family.

All models achieve perfect accuracy (100%) in the "No Functions Available" scenario, indicating reliable text-only response generation when no tools are present.

## 6.3 Error Analysis

The distribution of error types provides insight into specific failure modes and the effectiveness of the proposed fine-tuning approach.

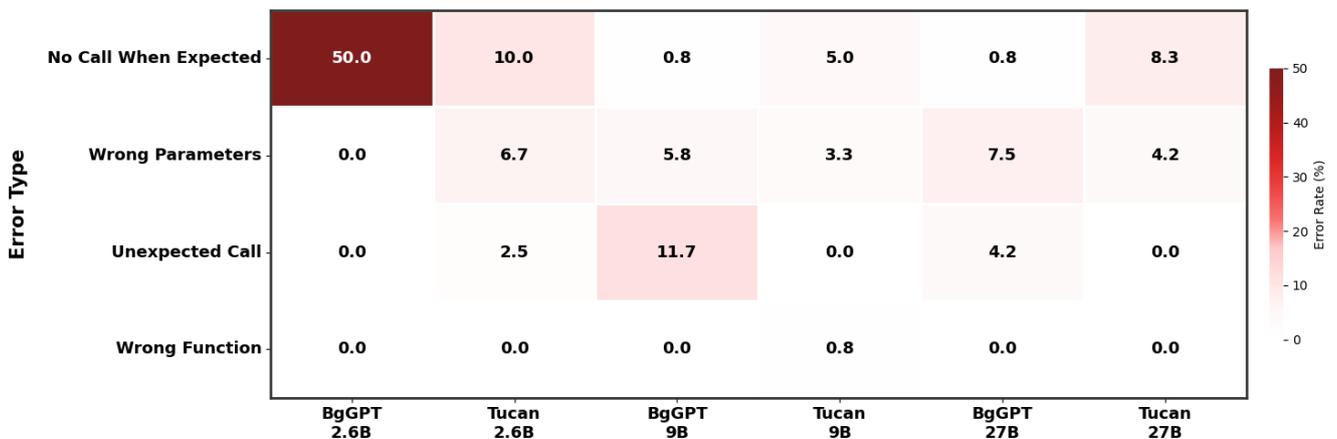

**Figure 6. Error distribution analysis**





The predominant error type in **BgGPT-2.6B** is **NO_CALL_WHEN_EXPECTED** (60 cases, 50.0% of all test cases), explaining its poor overall performance. TUCAN models significantly reduce this error type across all sizes, with **Tucan-2.6B** reducing it to 12 cases (10.0%).

**UNEXPECTED_CALL** errors are eliminated in **Tucan-9B** and **Tucan-27B** models, indicating improved discrimination between scenarios requiring tools versus those that do not. **BgGPT-9B** shows the highest frequency of this error type (14 cases, 11.7%).

**WRONG_PARAMETERS** represents the most common error type among successful TUCAN models, suggesting that while these models correctly identify function requirements and appropriate tools, parameter extraction remains challenging in complex scenarios.

Notably, no models produced **MALFORMED_JSON** errors, indicating that all models successfully learned the structural requirements of the function-calling format.

## 6.4 Knowledge Retention Assessment

To verify that function-calling specialization did not induce model damage, all models were evaluated on standard Bulgarian language benchmarks using the lm-evaluation-harness-bg framework. The multi-run evaluation protocol was applied consistently across all benchmarks.

**Table 6. Knowledge Retention on Bulgarian Benchmarks (Averaged Across Multiple Runs)**

| Model | HellaSwagBG | WinograndeBG | ARC-Easy-BG | ARC-Challenge-BG |
|---|---|---|---|---|
| BgGPT-2.6B-IT-v1.0 | 0.5871 | 0.6306 | 0.5657 | 0.3720 |
| Tucan-2.6B-v1.0 | 0.5924 | 0.6456 | 0.5657 | 0.3754 |
| BgGPT-9B-IT-v1.0 | 0.7057 | 0.7190 | 0.7231 | 0.5188 |
| Tucan-9B-v1.0 | 0.7046 | 0.7151 | 0.7024 | 0.5188 |
| BgGPT-27B-IT-v1.0 | 0.6200 | 0.6212 | 0.6587 | 0.4590 |
| Tucan-27B-v1.0 | 0.6179 | 0.6275 | 0.6486 | 0.4420 |

The results demonstrate that TUCAN models maintain performance comparable to their base models across all benchmarks. The maximum deviation is 0.0382 points on HellaSwagBG for the 2.6B model, while most differences are within 0.02 points. These minimal variations fall within expected measurement noise and do not indicate systematic performance degradation. **BgGPT-27B-IT-v1.0** and **Tucan-27B-v1.0** were tested with 8-bit quantization for performance comparison between the two main models (which explains their relatively lower scores). Standard errors from the lm-evaluation-harness framework range from ±0.0045 to ±0.0146 across benchmarks, indicating that the observed performance differences between TUCAN and base models are within typical measurement variability for these evaluation protocols.

**Tucan-2.6B** shows slight improvements on WinograndeBG (+0.0635) and ARC-Challenge-BG (+0.0034), with an average improvement of +0.0176 across all benchmarks. **Tucan-9B** and **Tucan-27B** show minor average decreases of -0.0050 and -0.0040, respectively, confirming that the LoRA fine-tuning approach successfully preserved the models' core linguistic competencies while adding function-calling capabilities.

## 6.5 Model Scaling Analysis

The relationship between model size and function-calling improvement reveals distinct scaling patterns across the two model families.





**BgGPT** models show strong positive scaling from 2.6B (50.00%) to 9B (78.33%) to 27B (86.67%), representing a 36.67 percentage point improvement from smallest to largest. **TUCAN** models demonstrate more compressed scaling: 2.6B (78.75%) to 9B (86.67%) to 27B (87.5%), representing an 8.75 percentage point improvement.

This compressed scaling pattern indicates that the fine-tuning process provides greater benefits to smaller models, effectively reducing the performance gap between different parameter scales for function-calling tasks. The relative improvement from fine-tuning decreases with model size: 28.75% for 2.6B, 8.34% for 9B, and 0.83% for 27B parameters.

The results establish that specialized fine-tuning for function-calling can achieve performance improvements while preserving core language understanding capabilities. The effectiveness varies inversely with model size, suggesting that larger base models already possess many of the reasoning capabilities required for tool use, while smaller models benefit more dramatically from targeted training.

## 6.6 Response Quality and Style Analysis

TUCAN models consistently generate clean, minimal function calls without extraneous text. For example, when prompted with "*Изчисли ми данъка върху имота с оценъчна стойност 150000 лв, жилищен тип, в община София*" (*Calculate the property tax for a residential property valued at 150,000 BGN in Sofia municipality)*, TUCAN models produce direct responses with proper JSON structure, correct parameter extraction, and absence of unnecessary explanatory text. This format demonstrates optimal characteristics for production deployment.

From the other end, **BgGPT** models generates functional but verbose responses, often including simulated function results and explanatory text before or after the function call. While this could be technically correct, this verbosity requires post-processing for production use.

**BgGPT-27B** demonstrates the most problematic response style, generating extensive explanatory text before and after function calls. For the same property tax query, BgGPT-27B produces responses such as: "*За да изчисля данъка върху имота, ще използвам функцията calculate_property_tax. Ето как изглежда извикването на функцията:*" (*To calculate the property tax, I will use the calculate_property_tax function. Here is how the function call looks:*)

This conversational approach, while potentially informative, creates parsing challenges and reduces efficiency in automated systems. Additionally, BgGPT-27B occasionally generates incorrect parameter structures, such as nesting arguments within unnecessary wrapper objects.

The response style differences have direct implications for user experience and system integration. TUCAN models provide the seamless, efficient interaction patterns expected in production AI systems. The verbose nature of larger BgGPT models, while potentially more "conversational" actually detracts from the functional utility by introducing unnecessary complexity and potential parsing errors.

## 6.7 Discussion and Broader Implications

While this work demonstrates successful adaptation of Bulgarian language models for function-calling capabilities, several methodological considerations warrant discussion for broader applicability and future research directions. The 120-case evaluation dataset, while carefully designed across six scenario types, represents controlled laboratory conditions rather than the complex, open-ended interactions characteristic of production deployments. Future work should incorporate larger-scale evaluation protocols and human





assessment metrics to validate practical utility in authentic user interactions. This paper establishes performance baselines against corresponding BgGPT models, though a comprehensive assessment would benefit from comparison against alternative approaches, including complex prompt engineering techniques and multilingual models with inherent tool-use capabilities.

## Conclusion

This work demonstrates a successful methodology for adapting multilingual language models to enable function-calling capabilities in non-English languages. Using Bulgarian as a case study, the BgGPT model series was fine-tuned on a novel bilingual dataset of 10,035 function-calling examples, introducing the TUCAN models that achieve improvements in tool-use accuracy: 28.75% for the 2.6B model, 8.34% for the 9B model, and 0.83% for the 27B model.

Critically, these improvements were achieved without compromising core language understanding, as verified on established Bulgarian benchmarks. The results show that smaller models benefit more dramatically from specialized function-calling training, effectively reducing the performance gap between different parameter scales for tool-use tasks.

Beyond raw performance gains, TUCAN models demonstrate production-ready response formatting with clean, parseable function calls, contrasting with the verbose and inconsistent outputs of base models. This combination of accuracy and practical utility positions TUCAN as optimal for real-world deployment.

The complete TUCAN model series, evaluation framework, and training dataset are released as open-source contributions. This work provides a replicable blueprint for extending tool-augmented capabilities to any language, helping bridge the existing gap between English-centric AI systems and global linguistic communities. The methodology can be directly applied to other languages to democratize access to advanced AI agent capabilities worldwide.

# Appendix A

**Note on Naming Convention:** TUCAN refers to the model series acronym (Tool-Using Capable Assistant Navigator), while repository and file names use the capitalized form "Tucan" for consistency with naming conventions.

**Table 4. Model releases**

| Model size | Full model | LoRA adapter | Quantized (GGUF) |
|:---:|:---:|:---:|:---:|
| 2.6B | Tucan-2.6B-v1.0 | Tucan-2.6B-v1.0-LoRA | Tucan-2.6B-v1.0-GGUF |
| 9B | Tucan-9B-v1.0 | Tucan-9B-v1.0-LoRA | Tucan-9B-v1.0-GGUF |
| 27B | Tucan-27B-v1.0 | Tucan-27B-v1.0-LoRA | Tucan-27B-v1.0-GGUF |

**Table 5. Additional resources**

| Resource type | Repository | Desctiption |
|:---:|:---:|:---:|
| Training Dataset | Tucan-BG-v1.0 | Bilingual function-calling training dataset (10,035 conversations) |
| Evaluation Dataset | Tucan-BG-Eval-v1.0 | Function-calling evaluation benchmark (120 test cases) |
| Evaluation Framework | Tucan-Eval | Open-source evaluation framework for function-calling assessment |

**Table 6. Deployment and collection resources**

| Resource type | Repository | Desctiption |
|:---:|:---:|:---:|
| Deployment Platform | Ollama Models | Easy deployment via Ollama platform |
| Complete Collection | Model Collection | Complete collection of all TUCAN models |

**Table 7. Base models (referenced)**

| Base model | Size | Purpose |
|:---:|:---:|:---:|
| BgGPT-Gemma-2-2.6B-IT-v1.0 | 2.6B | Base model for Tucan-2.6B |
| BgGPT-Gemma-2-9B-IT-v1.0 | 9B | Base model for Tucan-9B |
| BgGPT-Gemma-2-27B-IT-v1.0 | 27B | Base model for Tucan-27B |

**Table 8. Evaluation framework (referenced)**

| Framework | Purpose |
|:---:|:---:|
| lm-evaluation-harness-bg | Bulgarian language benchmark evaluation framework |

All contributions are released under open-source licenses to facilitate reproduction and extension to other languages.